\documentstyle[12pt]{article}
\begin{document}
\textheight 23.5cm \textwidth 16.5cm
\voffset -20mm \hoffset -5mm
\pagestyle{empty}
\title{BEAM DYNAMICS PROBLEMS IN A MUON COLLIDER}
\author{
Robert B. Palmer, Juan C. Gallardo\thanks{E-mail:jcg@fel.cap.bnl.gov},
Richard C. Fernow, Harold Kirk,\\
I. Stumer, Y. Y. Lee, M. Syphers, Ya\u{g}mur Torun\thanks{
Brookhaven National Laboratory,
P. O. Box 5000, Upton, New York  11973-5000}
\and
David Winn \thanks{
Fairfield University,
Fairfield, CT, 06430-5195}
\and
David Neuffer \thanks{
CEBAF,
Newport News, VA, 23606}
\and
Yang Cho, James Norem \thanks{
Argonne National Laboratory,
Argonne, Illinois 60439-4815}
\and
Nikolai Mokhov, Robert Noble, Alvin Tollestrup \thanks{
Fermi National Accelerator Laboratory,
 P.O. Box 500,  Batavia, Illinois 60510}
\and
Ronald Scanlan, Shlomo Caspi\thanks{
Lawrence Berkeley Laboratory,
Berkeley, CA 94720}
\and
Olivier Napoly \thanks{
DAPNIA-SEA, CE Saclay, 91191 Gif/Yvette CEDEX, France}}
\date{\today}
\maketitle
\begin{abstract}
We discuss the various beam dynamics problems in muon collider systems,
starting from the proton accelerator needed to generate the muon beams and
proceeding through the muon storage ring.

\phantom{We discuss the various beam dynamics problems in muon collider
systems,
starting from the proton accelerator needed to generate the muon beams and
proceeding through the muon storage ring.}

\vspace{3in}
\end{abstract}
\newpage

\section*{INTRODUCTION}

Lepton $(e^+e^-)$ colliders have the valuable property of producing simple,
single-particle interactions with little background, and this property is
essential in the exploration of new particle states. However, extension of
$e^+e^-$  colliders to multi-TeV energies is severely performance-constrained
by beamstrahlung, and cost-constrained because two full energy linacs are
required\cite{ref1}. On the other hand muons (heavy electrons) have negligible
beamstrahlung, and can be accelerated and  stored in rings.

The liabilities of $\mu$'s are that they decay, with a lifetime of  $2.2\times
10^{-6}$ s, and that they are created through decay into a diffuse phase space.
 In addition the decay products are likely to create large backgrounds at the
final focus points making the detector design a challenge. The first problem is
overcome by rapidly increasing the relativistic $\gamma$ factor; at 2 TeV for
example, the lifetime is $0.044\,$s, sufficient for storage-ring collisions.
The second can be dealt with by cooling. The possibility of muon colliders has
been introduced  by Skrinsky et al.\cite{ref2}, Neuffer\cite{ref3},  and
others. More recently, several workshops and collaboration meetings have
greatly increased the level of discussion\cite{ref4},\cite{ref5}.  In this
paper we discuss the beam dynamics problems encountered in one particular
scenario for a 2 + 2 TeV collider. Table 1 shows parameters for the candidate
design. This scenario includes a high-intensity $\mu$-source, $\mu$-cooling,
acceleration and storage in a collider. The complete cycle is repeated at
30 Hz.
\begin{table}[b]
\centering
\protect \caption{Summary of Parameters of 2 + 2 TeV Muon-Muon Collider}
\vspace{7 mm}
\begin{tabular}{|ll|c|}
\hline
Beam energy                & TeV      &     2    \\
Beam $\gamma$              &          &   19,000 \\
Repetition rate            & Hz       &    30   \\
Muons per bunch            & $10^{12}$  &   2     \\
Bunches of each sign       &          &   1     \\
Normalized rms emittance $\epsilon_n$   &mm mrad  &  50      \\
Average ring mag. field $B$    & Tesla   & 6      \\
Effective turns before decay &       & 900     \\
$\beta^*$ at intersection   & mm     &   3      \\
\smallskip
Luminosity  ${\cal L}$       &${\rm cm}^{-2}{\rm s}^{-1}$&  $10^{35}$\\
\hline
\end{tabular}
\label{sum}
\end{table}
\section*{SYSTEM COMPONENTS}
\subsection*{Proton Driver}

The $\mu$-source driver is a high-intensity rapid-cycling ($30\,$Hz) proton
synchrotron. A recent study\cite{ref6} suggests that an optimum proton energy
may be 10 GeV. In this case we require a total of about $10^{14}$ protons at
$30\,$Hz. This specification is almost identical to that studied\cite{ref7} at
ANL for a spallation neutron source. The only difference is the number of
bunches: 2 of $5\times 10^{13}$ instead of one of $10^{14}$. One of which is
for making  $\mu^-$, the other for $\mu^+$, both brought onto the same target.

A problem specific to muon colliders is that the proton bunches should be
short, with an rms bunch length less than 3 ns (1 m).

An RF sequence must be designed to phase rotate the bunch prior to extraction.
The total final momentum spread, based on the ANL parameters (95\% phase space
of $4.5\,{\rm V s}$  per bunch), is modest ($6\,\%,$ or $2.5\, \%$ rms), but
the space charge tune shift just  before extraction would be very large
$(\approx 1.5).$ A separate superconducting  compression ring is thus needed
(reducing the tune shift to $\approx 0.15$), or some  other  more exotic
solution must be found. Some possible parameteres of the main components of the
proton driver are given in table \ref{driver}.
\begin{table}[h]
\centering
\protect\caption{Proton Driver parameters}
\vspace{7 mm}
\begin{tabular}{|lll|c|}
\hline
RFQ     & Energy    &  MeV   &  2   \\
        & Frequency & MHz    & 400   \\
\hline
DTL     & Energy    &  MeV   & 68   \\
        & Frequency & MHz    & 400   \\
\hline
Linac   & Energy    &  MeV   & 330   \\
        & Gradient  & MeV/m  & 4-5   \\
        & Frequency & MHz    & 1200   \\
\hline
Booster 1 & Energy &  GeV    & 2.2   \\
          & Circ.  &  m      & 190   \\
          & Frequency & MHz  & 2.2-3 \\
\hline
Booster 2 & Energy &  GeV    & 10    \\
          & Circ.   &  m      & 690   \\
          & Frequency & MHz  & 9   \\
\hline
Buncher   & Energy   & GeV   & 10       \\
          &  Circ.    & m     & 70       \\
\hline
Final     & rms emittance&mm mrad&62 \\
          & rms long. phase space & V s & 0.7 \\
          & rms bunch length & ns & 3   \\
          & rms dp/p         & \%     & 2.5   \\
\hline
\end{tabular}
\label{driver}
\end{table}
\subsection*{Target and Pion Capture}

The target could be either copper (24 cm by 12 mm diameter) or beryllium (70 cm
by 2
cm diameter), although Cu would be preferred because of its higher pion
multiplicity. Pions are captured from the target by a high-field hybrid
solenoid that surrounds it. A field of 28 Tesla, and a radius of 7.5 cm are
consistent with what is currently available\cite{ref8}. The pions can then be
matched, using a suitable tapered field\cite{ref9} into a long (350 m)
solenoidal decay channel. A field of 7 Tesla and radius of 15 cm for the decay
channel seems reasonable and matches the capture acceptance.

Monte Carlo studies indicate that such a system captures almost 40\% of the
produced pions. Using the Wang\cite{ref10} formula for pion production, the
program calculates a yield of 0.22 muons, of each sign, per initial proton.
However, for a Cu target, a higher multiplicity is expected and consequently,
it would give a yet higher yield.

\subsection*{Phase Rotation Linac}

The pions, captured by a solenoidal focusing system (and the muons into which
pions
decay) have a huge energy spread, from 0 - 3 GeV (rms/mean $\approx 100 \%$),
and would be difficult to transport and to handle in any subsequent system. It
is thus proposed to introduce a linac along the decay channel, whose
frequencies and phases are chosen to deaccelerate the fast particles and
accelerate the slow ones; i.e. to phase rotate the muon bunch. Table \ref{rot}
gives the parameters of these linacs.
\begin{table}
\centering
\protect\caption{Parameters of Phase Rotation Linacs}
\vspace{7 mm}

\begin{tabular}{|c|cccc|}
\hline
Linac     & Length & Frequency & Gradient & Phase \\
          &  m     &   MHz     &  MeV/m   &  degrees\\
\hline
1         &  50    &   24     &   2     &   36 \\
2         &  50    &   24     &   2     &   0   \\
3         &  250   &   6      &   2     &   43  \\
4         &  60    &   24     &   2     &   81  \\
\hline
\end{tabular}
\label{rot}
\end{table}

After phase rotation  the rms bunch length is $6\,$m, and the rms momentum
spread is reduced to about $15\,$\%. Unfortunately, at such frequencies, the
linacs cannot phase rotate both signs in the same bunch: hence the need for two
bunches. The phases must be set to rotate the $\mu^+$'s of one bunch and the
$\mu^-$'s of the other.

\subsection*{Ionization Cooling}

\subsubsection*{Cooling Theory}

For collider intensities, the phase-space volume must be reduced within the
$\mu$ lifetime. Cooling by synchrotron radiation, conventional stochastic
cooling and conventional electron cooling are all too slow. Optical stochastic
cooling\cite{ref11}, electron cooling in an plasma discharge\cite{ref12} and
cooling in a crystal lattice\cite{ref13} are being studied, but are not by any
means certain. Ionization cooling of muons\cite{ref14} seems relatively
straightforward.

In ionization cooling, the beam loses both transverse and longitudinal momentum
as it passing through a material medium. Subsequently, the longitudinal
momentum can be restored by coherent reacceleration, leaving a net loss of
transverse momentum. Ionization cooling is not practical for protons and
electrons because of nuclear scattering (p's) and bremsstrahlung (e's) effects
in the material, but is practical for $\mu$'s because of their low nuclear
cross section and relatively low bremsstrahlung.

The equation for transverse cooling (with energies in GeV)  is:
  \begin{equation}
\frac{d\epsilon_n}{ds}\ =\ -\frac{dE_{\mu}}{ds}\ \frac{\epsilon_n}{E_{\mu}}\ +
\ \frac{\beta_{\perp} (0.014)^2}{2\ E_{\mu}m_{\mu}\ L_R}\label{eq1}
  \end{equation}
where $\epsilon_n$ is the normalized emittance, $\beta_{\perp}$ is the betatron
function at the absorber, $dE_{\mu}/ds$ is the energy loss, and $L_R$  is the
material radiation length.  The first term in this equation is the coherent
cooling term and the second term is the heating due to multiple scattering.
This heating term is minimized if $\beta_{\perp}$ is small (strong-focusing)
and $L_R$ is large (a low-Z absorber).

{}From Eq.\ref{eq1} we find a limit to transverse cooling, which occurs when
heating due to
multiple scattering  balances cooling due to energy loss. The limits are
$\epsilon_n\approx\ 0.6\ 10^{-2}\ \beta_{\perp}$ for Li, and
$\epsilon_n\approx\ 0.8\  10^{-2}\ \beta_{\perp}$ for Be.

The equation for energy or longitudinal cooling is:
 \begin{equation}
{\frac{d(\Delta E)^2}{ds}}\ =\
-2\ {\frac{d\left( {\frac{dE_\mu}{ds}} \right)} {dE_\mu}}
\ <(\Delta E_{\mu})^2 >\ +\
{\frac{d(\Delta E_{\mu})^2_{straggling}}{ds}}\label{eq2}
 \end{equation}
Where the first term is the cooling (or heating) due to energy loss
and the second term is the heating due to straggling.

Cooling requires that  ${d(dE_{\mu}/ds)\over dE_{\mu}} > 0.$ But at energies
below about 200 MeV, the energy loss function for muons, $dE_\mu/ds$, is
rapidly decreasing with energy and there is thus rapid heating of the beam.
Above 400 MeV the energy loss function increases gently, thus giving some
cooling, though not sufficient for our application.

In the long-path-length Gaussian-distribution limit, the heating term (energy
straggling)  is given by\cite{ref15}
 \begin{equation}
\frac{d(\Delta E_{\mu})^2_{straggling}}{ds}\ =\
4\pi\ (r_em_ec^2)^2\ N_o\ \frac{Z}{A}\ \rho\gamma^2\left(1-
\frac{\beta^2}{2}\right)
 \end{equation}
where $N_o$ is Avogadro's number and $\rho$ is the density. Since the energy
straggling increases as $\gamma^2$, and the cooling system size scales as
$\gamma$, cooling at low energies is desired.

Energy spread can also be reduced by artificially increasing
${d(dE_\mu/ds)\over dE_{\mu}}$ by placing a transverse variation in absorber
density at a location where position is energy dependent, i.e. where there is
dispersion. The use of such wedges can reduce energy spread, but it
simultaneously increases transverse emittance in the direction of the
dispersion. Six dimensional phase space is not reduced. But it does allow the
exchange of emittance between the energy and transverse directions.

\subsubsection*{Cooling System}

We require a reduction of the normalize transverse emittance by almost three
orders of magnitude (from $2\times 10^{-2}$ to $3\times 10^{-5}\,$m-rad), and a
reduction of the longitudinal emittance by more than an order of magnitude.
This cooling is obtained in a series of cooling cells. Each cell consists of a
section of beryllium $(\approx 0.7 \,m)$ or lithium $(\approx 2\,m)$ placed in
a region of the lattice with a low $\beta_{\perp}$, a linac $(200\,MeV),$ and a
matching bend with dispersion where wedges can be introduced to interchange
longitudinal and transverse emittance. The energy would be restricted to
between 200 and 400 MeV, so as to avoid the energy dE/dx heating below 200 MeV,
but minimize the straggling heating at higher momenta. About 20 such cells
would be needed.

For the early cells, when the emittance is still large, a sufficiently low
$\beta_{\perp}$ can be obtained with solenoids. In later cells, when the
emittance is lower and a lower $\beta_{\perp}$ is required, current carrying
cooling rods (approx 2 m long, if Li) which serve both to maintain the low
$\beta_{\perp}$ and reduce the beam energy could be employed. In a lithium rod,
with surface fields of 10 Tesla (as achieved in lithium lenses at Novosibirsk,
FNAL and CERN \cite{ref16}), a $\beta_{\perp}$ of 1.7 cm can be achieved, and
the emittance is reduced to about $10^{-4}\,$m. But  this is still a factor of
$\approx 3$ above the emittance goal of Table 1. A final stage might consist of
short sections of Be at even lower $\beta_{\perp}$ insertions. Alternatively,
the additional transverse emittance reduction can be obtained by cooling more
than necessary longitudinally, and then exchanging transverse and longitudinal
phase-space with a thick wedge absorber.

In all these cells lattices are required with adequate momentum acceptance,
matching in and out of the low beta insertions, appropriate momentum compaction
and control of emittance growth from space charge, wake field and resistive
wall effects. In addition it would be desirable to economize on linac sections
by forming groups of cells into recirculating loops.

\subsection*{Acceleration}

Following cooling and initial bunch compression (to of the order of $0.2\,$m)
the beams must be accelerated to full energy (2 TeV). A single linac of this
energy would work, but would be expensive, and would not utilize our ability to
recirculate $\mu$'s in rings. A conventional synchrotron cannot be used because
the muons would decay before they were accelerated. A fast cycling synchrotron
could be used but, because it would be limited to low magnetic fields, would be
very large. The best solution seems to be a recirculating linac (similar to
CEBAF). If acceleration is done in 20 recirculations, then only 100 GeV of
linear accelerator is required.

In practice,  a cascade of at least 3 recirculating linacs (e.g., with maximum
energies of 20 GeV, 200 GeV and 0.2 TeV) would be needed. The $\mu$-bunches
would be compressed  on each of the return arcs, and be bunched finally to the
required length of 3 mm at full energy. The two higher energy recirculators
must be superconducting for two reasons: the store time is far too long for
conventional cavities, and the wall power consumption with conventional
cavities would be too high. The total muon beam power is 38 MW. It is hoped to
achieve at least 30\% efficiency with superconducting cavities, giving a wall
power consumption of 127 MW. The gradients assumed are below those assumed for
TESLA. They may be over conservative in view of the shorter pulse duration in
this application than assumed in TESLA. The muon linac beam dynamics is
complicated by transverse HOM because of the large number of muons per bunch,
about a factor of 100 higher than electrons in TESLA. The HOM power is
estimated to be $\approx 100\,$W/m. As in the TESLA design, this would required
a coupler section to remove this HOM power.

At the higher energies, space charge effects will not be a problem, but as the
bunches are compressed wake field and resistive wall effects become serious.
Preliminary studies suggest that, with a slight decrease in Q/Z (by widening
the irises, and with BNS damping, such effects can be controlled.

\subsection*{$\mu$ Storage Ring}

After acceleration, the $\mu^+$ and $\mu^-$ bunches are injected into the 2-TeV
storage ring ,  with collisions in two low-$\beta^*$ interaction areas. The
beam size at collision is $r=\sqrt{\epsilon_n\beta^*} \approx 2\,\mu m,$
similar to hadron collider values. The bunch populations decay exponentially,
yielding an integrated luminosity equal to its initial value multiplied by an
``effective'' number of turns $n_{effective}=150\ B,$ where B is the mean
bending field in T. With 9 T superconducting magnets, an average B of 6 Tesla
might be obtained, yielding an $n_{effective}\approx 900.$ The magnet design is
complicated by the fact that the $\mu$'s decay within the rings ($\mu\
\rightarrow\ e\nu_e\nu_{\mu}$), producing electrons whose mean energy is
approximately 1/3 of that of the muons. These electrons travel to the inside of
the ring dipoles, but radiate a substantial fraction of their energy, as
synchrotron radiation, towards the outside of the ring. A warm tungsten liner
of about 2 cm thickness will be required to intercept this radiation.

Even a preliminary study of  resistive wall impedance instabilities indicate
that 3 mm bunches of $2\times 10^{12}$ muons would be unstable in a
conventional ring. In any case, the rf requirements to maintain such a bunch
are excessive. It is thus proposed to use an isochronous lattice of the type
discussed by S.Y. Lee et al\cite{ref17}. It remains to be seen if the required
high degree of isochronism can in fact be achieved.

Another problem is the design of chromatic correction for the very low beta
($\beta^*=3\,mm$) insertions. A triplet design would have maximum beta's of 400
km in both directions, and chromaticity $(1/4\pi\int \beta dk)$ of 3,700 in x
and 3,500 in y. It seems clear that a local correction of
chromaticity\cite{ref18} would be required. A preliminary
automated\cite{ref18a}  study of such a correction system, using a doublet at
the final focus, gave momentum acceptances of $\pm 0.1\,\%$ and $\pm 0.6\,\%$
in the two directions, where the $\beta_{max}$'s were 1.2 and 0.2 million m
respectively. A similar design with the triplet ($\beta_{max}$'s both 0.4
million m) would be expected to give about $0.3\,\%$ in both directions. More
sophisticated designs \cite{ref19} should do better. But this estimate is only
for a single pass device like a linear collider; the performance for a storage
ring remains to be seen.

\subsubsection*{Detector Background}

For the physics user there is a problem of background from $\mu$-decays that
occur through the whole collider ring, from near the interaction region and
from scattering of any muon halo circulating in the ring.

A first Monte Carlo study\cite{ref20} has been done with the MARS95
code\cite{ref21}, based on a preliminary insertion lattice. A contribution of
direct decays has been studied in detail and a few ways to supress the
background levels in a generic detector have been proposed. Also, from this
study it is clear that much attention should be paid to muon halo contribution.
A collimation system will be required in a straight section far from the
detectors (presumably a quarter way around the ring). No such system has yet
been designed.

Background track densities initiated by muon decays are indicated in Table
\ref{back}. In this study it was assumed that the detector pixels in the inner
tracker were 20 $\mu m$ by 20 $\mu m$, and in the central tracker: 50 $\mu m$
by 300 $\mu m$. The track densities are high, but they result from very low
energy electrons that would be eliminated in any track reconstruction. Given
the fine subdivision of the assumed detector, the occupancies do not look
impossible. However, it is hoped, and expected, that the background can be
greatly reduced from these values by further improvements in shielding.
\begin{table}
\centering
\protect\caption{Detector Backgrounds from $\mu$ decay}
\vspace{7 mm}

\begin{tabular}{|l|cc|cc|}
\hline
Location   & \multicolumn{2}{c|}{outside}&\multicolumn{2}{c|}{inside}\\
\hline
           &  density   & occupancy  &density  & occupancy \\
               & $cm^{-2}$&  \%      & $cm^{-2}$   \% &    \\
\hline
inner tracker  &  170    &   0.07  &  480    &     0.19    \\
central tracker&   3.2   & 0.05    &   2.3   &    0.03   \\
outer tracker  &   1.7   &  -      &    0.3  &    -    \\
\hline
\end{tabular}
\label{back}
\end{table}
\section*{CONCLUSION}

\begin{itemize}
\item The scenario for a 2 + 2 TeV, high luminosity collider is by no means
complete. There are many problems still to be examined. Much work remains to be
done, but:
\item No obvious show stopper has yet been found.
\item Many technical components require development: a large high field
solenoid for capture, low frequency rf linacs, long lithium lenses, multi-beam
magnets for recirculation, warm bore shielding inside high field dipoles for
the collider, muon collimators and background shields, but:
\item None of the required components may be described as {\it exotic} and
their specifications are not far beyond what has been demonstrated.
\item If the problems can be overcome, then a muon-muon colliders may be the
best route to study physics at energies higher than those accessible at the LHC
or NLC.

\end{itemize}

\section*{ACKNOWLEDGMENTS}

We acknowledge extremely important contributions from our
colleagues, especially  C. Pellegrini, D. Cline, A. Chao, A.
Ruggiero, A. Sessler, J. D. Bjorken, W. Barletta, D.Douglas, F. Mills, W.
Willis, Dick Helms, Y. Zhao, H. Padamsee, John Irwin and Z. Parsa.

\medskip
This research was supported by the U.S. Department of Energy under Contract No.
DE-ACO2-76-CH00016 and DE-AC03-76SF00515.

\end{document}